\documentclass{article}
\usepackage{fullpage}

\usepackage{amsmath, amssymb, graphicx, multirow, epsfig, subfigure}
\usepackage{cite, bbold}

\def\argmax{\operatornamewithlimits{arg\,max}}

\def\a{\alpha}

\newtheorem{definition}{Definition}

\newtheorem{observation}{Observation}

\title{Sponsored Search Auctions with Rich Ads}

\author{Ruggiero Cavallo\footnote{Yahoo Research, New York, NY. \texttt{cavallo@yahoo-inc.com}.} \and
Prabhakar Krishnamurthy\footnote{Yahoo Research, Sunnyvale, CA. \texttt{pkmurthy@yahoo-inc.com}.} \and
Maxim Sviridenko\footnote{Yahoo Research, New York, NY. \texttt{sviri@yahoo-inc.com}.} \and
Christopher A. Wilkens\footnote{Yahoo Research, Sunnyvale, CA. \texttt{cwilkens@yahoo-inc.com}.}}

\begin{document}
\maketitle

\begin{abstract}
The generalized second price (GSP) auction has served as the core selling
mechanism for sponsored search ads for over a decade. However, recent trends
expanding the set of allowed ad formats---to include a variety of sizes,
decorations, and other distinguishing features---have raised critical problems
for GSP-based platforms. Alternatives such as the Vickrey-Clarke-Groves (VCG)
auction raise different complications because they fundamentally change the way
prices are computed. In this paper we report on our efforts to redesign a
search ad selling system from the ground up in this new context, proposing a
mechanism that optimizes an entire slate of ads globally and computes prices
that achieve properties analogous to those held by GSP in the original, simpler
setting of uniform ads. A careful algorithmic coupling of
allocation-optimization and pricing-computation allows our auction to operate
within the strict timing constraints inherent in real-time ad auctions. We
report performance results of the auction in Yahoo's Gemini Search platform.
\end{abstract}

\sloppy
\section{Introduction} \label{sec:intro}

Very early in the history of sponsored-search advertising, all the major
platforms settled on some version of the Generalized Second Price (GSP) auction
as the mechanism used to sell search ad spots. GSP has had remarkable staying
power, apparently serving search ad marketplaces well for over a decade. 
However, recent trends expose problems stemming from the rigidity of
traditional GSP-bound platforms: ads now come in various sizes and formats, and
a mechanism that simply sorts ads and prices each based on competition from the
ad below will have significant inefficiencies and unsought incentive
properties.

\begin{figure}[h!] 
\center \includegraphics[scale=0.95]{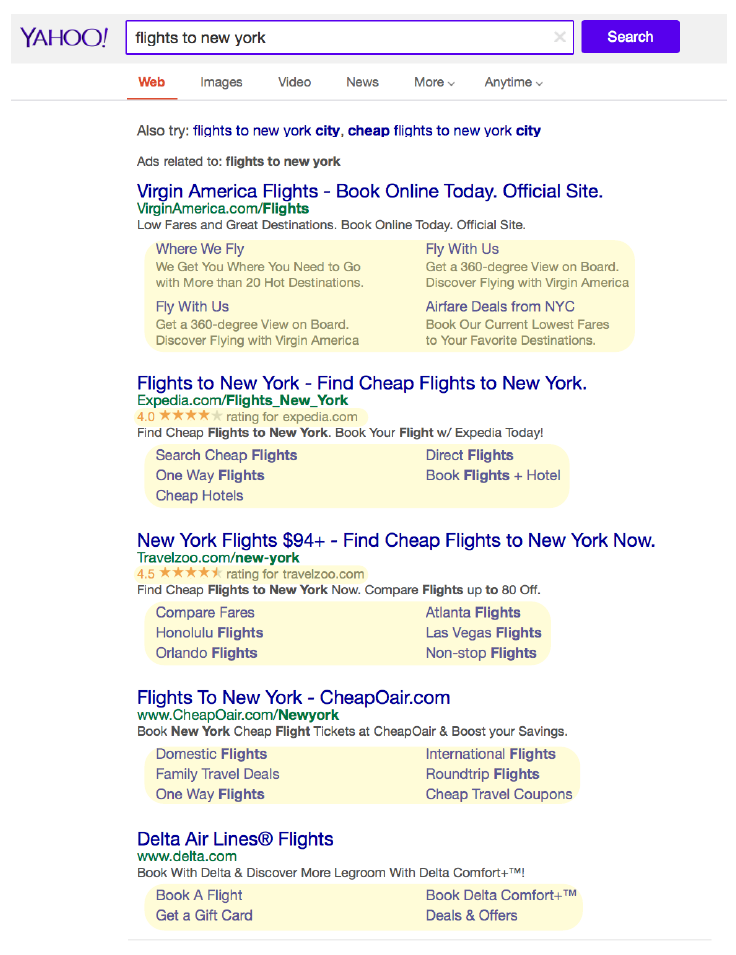}
\caption{\label{fig:adex} All highlighted sections of the above image are optional extensions to the basic three-line ad format. The search platform can include or omit them at its discretion in order to optimize the overall slate of ads presented to the user.}
\end{figure}

For instance, imagine that a search platform has a priori allotted 12 lines at
the top of the search page for advertisements. In the ``old world'' all ads
were three lines long (just a title, url, and description), and so in this
12-line example there would be precisely four available ad {\em slots},
regardless of which advertisers bid. But in the ``new world'' there may be ads
that have the basic three lines {\em plus} additional lines of sitelinks
(taking the user directly to specific sections of the advertiser's landing
page), star-ratings, location information, a phone number, etc. An example of
some of these ad extensions and decorations on Yahoo's search platform is given
in Figure \ref{fig:adex}.

\begin{figure}[h] 
\center \includegraphics[scale=1.15]{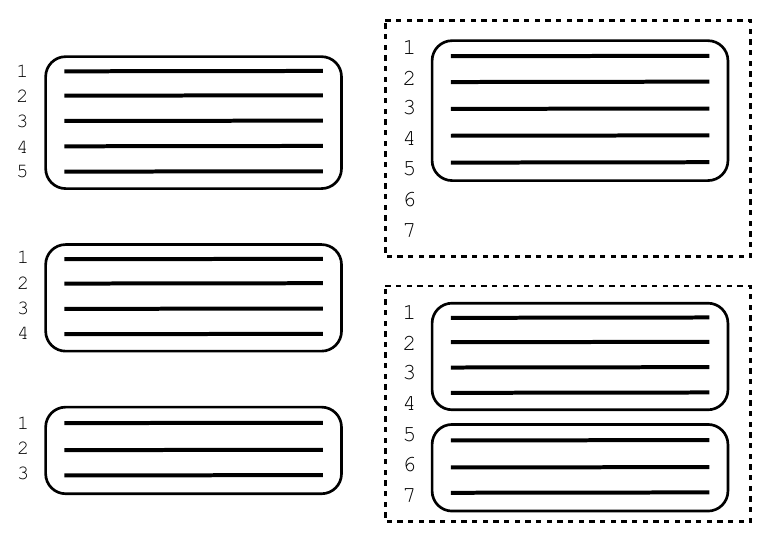} 
\caption{\label{fig:pack} An illustration of the ad packing problem that arises in a context with ads of varying size. Imagine there are three ad candidates, of size 5, 4, and 3, respectively (illustrated at left), and seven lines of total available ad space. Imagine that a ranking of the ads (by whatever metric one chooses) puts them in decreasing order of size. A greedy algorithm will place the five-line ad, and then have no ability to fill the remaining two lines (top right). It may be more efficient to instead place the four- and three-line ads (bottom right).}
\end{figure}

In the new world of heterogeneous ads, an {\em ad packing problem} emerges.
In a context where ads can vary in length, the search platform faces the richer
problem of deciding which versions of which ads---and how many---to show.
Whether it's best to show a larger or smaller version of an ad may depend on
which size-variants of competing ads are available.  Perhaps one giant ad
should be chosen to fill the entire space, or perhaps it's better for the giant
ad to be ``trimmed'' to a more moderate size and paired with a second small ad
below it, or perhaps a slate of several three-line ads is best, etc.
An illustration of the packing problem is given in Figure \ref{fig:pack}.

To deal with this new world, in this paper we rethink the search ad allocation
and pricing problems from the ground up, proposing a mechanism that optimizes
an entire page of ads {\em globally}. The efficiency-maximizing ad allocation
problem can be formulated as an integer program; however, for a number of
reasons, solving it this way is unwieldy in practice. We instead approach the
problem explicitly as a search through the space of possible ad slates;
the specific solution we implement is local-search based and not guaranteed to
find an optimal configuration, but in practice its distance from optimality is
negligible.

The main feature of the classic approach that we retain is a separable
click-probability model, wherein it is assumed that the probability that a
given ad will be clicked is equal to the product of an ad-variant-specific ``ad
clickability'' number and an ad-variant-independent ``location clickability''
number. Even here, though, innovations are required: the ``location
clickability'' number can no longer be associated with an ad {\em slot}, since
the starting-position of the $i^{th}$ ad now depends on what kind of ad
variants were shown above---location clickability for us is now a function of
starting-line-position rather than slot-number.

The most technically novel contribution of this work is probably the approach
we take to computing prices. It is most common to think of an allocation and
pricing mechanism as proceeding in two stages---an allocation is computed, and
from it (and the bids that generated it) prices are subsequently calculated.
This two-phase approach makes a great deal of conceptual sense, but in our
setting even fractions of a millisecond of compute-time are critical, and so a
more integrated solution was required. Noting that the most salient pricing
schemes can all be described in terms of the {\em allocation function} $x$
(where $x_i(b_i)$ is the probability of bidder $i$ receiving a good---here, an
ad click---given that he bids $b_i$), during the local search phase of the
algorithm we ``log'' key information from which we can, ultimately, very
quickly compute each bidder's allocation function. Then, whatever pricing
scheme is chosen, it can be applied by essentially just ``reading off'' prices
from the computed allocation functions.

\subsection{Related work}  \label{sec:rw}

The problem of rich ads in search is well known, but not as well studied. In
one sense, there is little to do --- the elegant Vickrey-Clarke-Groves (VCG)
auction reduces any problem related to rich ads to a modeling and optimization
problem if one buys into it, and Facebook and Google have both leveraged VCG
for this very reason~\cite{varian14}.

The primary challenge for rich search ads is that marketplaces have been
running GSP auctions for years; two thin lines of work consider the
consequences. The first line of work studies GSP-like mechanisms in more
complex optimization domains: Papers by Deng et al.~\cite{deng10}, Bacharach et
al.~\cite{bachrach14}, and Cavallo and Wilkens~\cite{cavallo14} generally show
negative results that equilibria may be poor or nonexistent. The second line of
work instead aims to convert GSP bidders to VCG bidders with minimal
issues~\cite{bachrach15}.

More broadly, a long line of work starting with Varian~\cite{varian07} and
Edelman, Ostrovsky and Schwarz~\cite{eos07} studies GSP and attempts to
rationalize its use, e.g., by showing the existence of good equilibria or
showing that GSP is more robust when click-through-rates have
error~\cite{milgrom10, dutting16}. More recently, we argue that advertisers do
not have quasilinear utilities and that GSP may in fact be the truthful
auction~\cite{cavallo15,wilkens16}.

These prior works are all largely theoretical in nature. In the current paper,
while we do make some conceptual and modeling contributions, a large emphasis
is on reporting about what we think is an interesting large-scale engineering
task: how to solve a computationally hard market-based problem in a feasible
amount of time under severe runtime constraints. That dimension of our work
strongly connects with many other studies from very different domains, such as
\cite{fujishima1999taming, gunluk2005branch, lubin08}, to name a few.

\section{The ad allocation problem}  \label{sec:allocation}

We start with a formal description of the ad allocation problem. There is a set
$A$ of ad ``candidates''; each $a \in A$ has a height $h(a)$ and is associated
with an advertiser $\a(a)\in \Lambda$, where $\Lambda$ is the set of all
advertisers. At most one ad per advertiser can appear on the page. There are
also configurable limits on the number and cumulative height of ads that can be
shown on a single page: no more than \texttt{ADLIM} ads occupying a total of
\texttt{H} lines can be selected.
 
Each ad $a$ has an associated bid $b_{\a(a)}$ and click probability density
$p_a$. $b_{\a(a)}$ can be interpreted as the advertiser's claim about how much
value he will receive should one of his ads be clicked (it is the same for all
of his ads). The click probability density $p_a$ is a more novel concept: it
can be thought of as a kind of normalized ``click probability per line'' for
the ad. In combination with the line-specific location-clickability
parameters,\footnote{These may be calculated naively based on empirical
click-through-rates for every line of the page. More sophisticated approaches
that seek to avoid selection bias may also be applied; we do not delve into
such details here.} it determines $p_a(k)$ for each $k \in \{ 0, \dots,
\texttt{H}-1-h(a)\}$; $p_a(k)$ is the search platform's estimate of the
probability with which $a$ will be clicked if it is placed at starting line
$i$.

Each ad also has an associated vector of ``costs'', where cost can loosely be
thought of as the externality the ad imposes (on the user, the search platform,
etc.) if it is shown; $c_a(k)$ denotes this cost when ad $a$ is placed at
starting line $k \in \{ 0,\dots,\texttt{H}-1-h(a)\}$. One simple way costs may
be deployed in practice is to assign a constant value to every $c_a(k)$, for
every $a$ and $k$, using that constant as a knob with which to tune the average
ad footprint.

For each ad $a \in A$ and line $j \in \{ 0, \ldots, \texttt{H}-1-h(a) \}$, let
${\cal L}_{aj}=\{k:  j-h(a)+1\le k\le j\}$, i.e., if an ad $a$ starts on line
$i\in {\cal L}_{aj}$ then the ad $a$ covers line $j$.

Our goal is to maximize efficiency, i.e., total advertiser value net of costs
imposed by the chosen configuration of ads. Letting $z_{a,k}$ be a boolean
variable denoting whether or not ad $a$ is placed at starting line $k$, we can
formulate the problem as follows:
\begin{align}
&\text{maximize}  && \displaystyle\sum_{k=0}^{\texttt{H}-1}\sum\limits_{a \in A} \left(b_a \, p_a(k) - c_a(k)\right)\cdot z_{a,k} \label{eq:obj} \\
&\text{subject to} && \displaystyle \displaystyle\sum_{k=0}^{\texttt{H}-1}\sum\limits_{a: \a(a)=i}  z_{a,k} \leq 1, \quad 
\forall i \in \Lambda,\label{ad_constraint}\\
&      && \displaystyle\sum\limits_{a\in A}  \sum_{k \in {\cal L}_{aj} }z_{a,k} \leq 1, \quad 
\forall j \in \{0, \ldots, \texttt{H}-1 \},\label{line_constraint}\\ 
&	 && \displaystyle \displaystyle\sum_{k=0}^{\texttt{H}-1}\sum\limits_{a \in A}   z_{a,k} \leq \texttt{ADLIM}, \label{total_constraint}\\
&	 && \displaystyle z_{a,k}\in \{0,1\} .
\end{align}

Constraint (\ref{ad_constraint}) says that we can choose at most one ad variant
per advertiser. Constraint (\ref{line_constraint}) says that each line can be
covered by at most one ad. This constraint also implicitly encodes the fact
that our solution can use at most $\texttt{H}$ lines.  Constraint
(\ref{total_constraint}) limits the total number of ads chosen by the solution.

The above is an integer program that can be solved with standard methods. Even
though the problem is strongly NP-hard (by the reduction from 3-PARTITION),
 the number of possible ad candidates is bounded, and so asymptotic
runtime analysis is really not relevant. However, the runtime constraints of
this environment are extremely severe---to create an experience of ``instant
service'' for search users, every millisecond counts, and there may not always
be time to solve this integer program.

\subsection{Our algorithm}  \label{sec:algo}

Motivated by runtime constraints, we opt for a local-search based heuristic
approach to the problem. Our algorithm, described below in Figure
\ref{fig:algo}, virtually always obtains an optimal solution, but in a much
shorter period of time; moreover, it has an ``anytime'' property --- in the
rare event of an instance that cannot be solved within our time-constraints,
the local search can be shut down and the intermediate solution taken.

The algorithm starts by doing something akin to traditional GSP: it orders ads
by bid times click probability---except here we use click probability {\em
density} since ads vary in length---and then chooses a slate greedily. But
while this is where traditional GSP ends, it is only a starting point for us.
The core of the algorithm iteratively modifies the slate through a series of ad
swaps until no improving swaps can be made. We find solutions in this way for
every possible size slate, and then choose the best one.

\begin{figure}[h!]
\center{ \fbox{ \begin{minipage}{0.9\textwidth}
For ad slate cardinality $K \in \{1, \ldots , \texttt{ADLIM}\}$:
\begin{enumerate}
\item[1.] A \textbf{greedy} starting allocation --
\begin{enumerate}
\item[(i)] Order $A$ by bid times click-probability density.
\item[(ii)] Select the first $K$ ad candidates in the ordered list, iteratively reducing the set of available candidates to respect the one-ad-per-advertiser constraint (Eq.~\ref{ad_constraint}).
\end{enumerate}

\item[2.] A \textbf{local search} loop of 1-for-1 ad swaps -- \vspace{1mm} \\
Observe objective value $X$ (Eq.~\ref{eq:obj}).  \vspace{1mm} \\ For each ad
$a$ in the current solution (from top to bottom):
\begin{enumerate}
\item[(i)] Remove $a$ from the slate.
\item[(ii)] For each ad $b$ in the set of ads that are not part of the current
solution (including ad $a$), for each slot that $b$ can feasibly be inserted
into:
\begin{itemize}
\item Insert $b$ and observe the objective value.
\item If it exceeds $X$, log the swap and go to (2.).
\end{itemize}
\item[(iii)] No swap for $a$ improved the objective, so return $a$ back to its
original position.  \end{enumerate}
\item[] Execution for the cardinality $K$ iteration completes when there exists
no 1-for-1 ad swap that improves the objective value.
\end{enumerate}
The best of the $K$ locally optimal solutions is chosen.
\end{minipage}}}
\caption{\label{fig:algo} A description of our heuristic ad allocation
algorithm. In the first phase a rank-based configuration, akin to what vanilla
GSP would produce, is chosen. Then in phase two it is iteratively improved
until a local optimum is reached.}
\end{figure}


What are the possible vulnerabilities of this algorithm---i.e., in what cases
might we get stuck in a local optimum that is not globally optimal?  This may
happen only in cases where swapping more than one ad at a time is required.
Note that the absence of ``1-for-2'' swaps and the like is strongly mitigated
by the fact that we find a local optimum for every possible cardinality ad
slate.  We will report detailed performance statistics in Section
\ref{sec:res}. For now, suffice it to say that the algorithm rarely leaves
significant efficiency on the table.

\section{Pricing}  \label{sec:pricing}

Our pricing implementation maximizes flexibility by estimating each bidder
$i$'s allocation curve $x_i$.\footnote{In our setting there are a variety of
non-null outcomes (ranging over ad variants and the slots they may appear in)
that any given bidder may receive; but an ``allocation'' can be reduced to the
one dimension that determines advertiser value: probability of click.
$x_i(b_i)$ is thus the probability with which $i$ receives a click in the
outcome yielded when he bids $b_i$ and all other bidders' bids are held
constant.} The allocation $x_i$ is a common tool in theory because it fully
captures what an advertiser needs to know when selecting a bid.  However,
auctions in practice rarely construct $x_i$ explicitly; instead, they rely on
computations that indirectly reference it. For example, externality pricing in
the VCG auction is computed by removing each bidder one at a time and computing
the negative effect on others --- this computation happens to be equal to the
area above $x_i$.

In our case, having direct access to $x_i$ is important for two reasons. First
and foremost, as we will discuss later, we strive to maintain GSP-like pricing,
and our formulation effectively requires full knowledge of the curve $x_i$.
Second, having access to $x_i$ gives substantial flexibility in pricing if
Yahoo wishes to change in the future, say, if competitors switch to a different
pricing function such as VCG and Yahoo feels compelled to follow suit.

We will first discuss how we estimate $x_i$ efficiently; then we will discuss a
handful of possible pricing strategies and motivate GSP-like prices.

\subsection{Estimating allocation curves}

\def\conf{\mathcal C}

The local search optimization explores a wide variety of slates; we
want to use these slates to efficiently construct an approximation of
$x_i$. Since the allocation curve will be piecewise-constant, our desired
output is a sequence of thresholds $\tau_i[0],\dots,\tau_i[k]$ and a sequence
of allocations $\hat x_i[0],\dots,\hat x_i[k]$, where the final estimated
allocation is given by:
\[\hat x_i(b_i)=\begin{cases}\hat x_i[j]&\tau_i[j-1]\leq b_i <\tau_i[j]\\ \hat
x_i[k]&b_i\geq \tau_i[k-1]\end{cases}\enspace.\]

This is conceptually easy to compute in a na\"{i}ve way: identify the
breakpoints $\tau_i$ by repeated binary search. Unfortunately, this will
require too much time, as the allocation algorithm must be run at every stage
of the binary search. We must therefore leverage the work of local search to
construct an approximation.

\paragraph{The approximate allocation $\hat x_i$}  \vspace{1mm}
Note that if we run the optimal algorithm, bidder $i$'s allocation can be
written as $x_i(b)=x_i(\argmax_\conf OBJ(\conf))$ where $x_i(\conf)$ denotes
the allocation probability (probability of a click) on $i$'s ad in
slate $\conf$.  Given any subset of possible slates
$S$, we can then define an approximation $\hat x_i$ by taking the $\argmax$
over only those slates in $S$, i.e., $\hat x_i(b)=x_i(\argmax_{\conf\in
S} OBJ(\conf))$.  We use this idea to define $\hat x_i$:

\begin{definition}
The {\em local search approximation} of the allocation curve is \[\hat
x_i(b)=x_i\left(\argmax_{\conf\in LS}OBJ(\conf)\right)\] where $LS$ denotes the
set of slates considered by the local search algorithm.
\end{definition}

Note that the approximation $\hat x_i$ is the exact allocation assuming that
the mechanism always explored a fixed set of slates and selected the
optimal one.\footnote{Said in terms of another standard mechanism, $\hat x_i$
is the allocation of a maximal-in-range allocation on the set of slates
$LS$.} However, since the mechanism will explore different slates for
different bids, $\hat x_i$ can both over- and under-estimate $x_i$. The
accuracy of $\hat{x}$ as an approximation of $x$ will be discussed in Section
\ref{sec:res}.

\paragraph{Computing $\hat x_i$ efficiently} \vspace{1mm}
Note that for any slate $\conf$, we have
\[  OBJ(\conf) = \sum_{i \in \Lambda} \Big( x_i(\conf) b_i - c_i(\conf) \Big)
\]
In particular, fixing bidder $i$ this can be written as $OBJ(\conf)=z_{i,\conf}
+ x_i(\conf) b_i$, where
$z_i,C = \sum_{j \in \Lambda \setminus\{i\}} \big( x_j(\conf) b_j - c_j(\conf)
\big) - c_i(\conf)$
is independent of $b_i$. If we let $\phi_i(b_i)$ denote the optimal objective
value when $i$ reports $b_i$ (holding $b_{-i}$ fixed), we can write:
\[ \phi_i(b_i)=\max_{\conf} \big\{ z_{i,\conf} + x_i(\conf) b_i \big\}\]
Each slate $\conf$ yields a distinct $z_{i,\conf}+x_i(\conf) b_i$
(i.e., objective value as a function of $b_i$) line, and $\phi_i$ is the
upper-envelope of these lines. Importantly, $i$'s allocation when reporting
$b_i$ is the slope of the upper envelope at $b_i$:

\begin{observation}
The optimal objective value $\phi_i$, as a function of bid $b_i$, for a set of
slates $S$ is the upper-envelope of the lines
$\{z_{i,\conf}+x_i(\conf)b_i\}$ associated with the slates $\conf \in
S$. The associated allocation function $\hat x_i$ is the slope of the upper
envelope $\frac{d\phi_i}{db_i}$.
\end{observation}

\begin{figure}[h!]
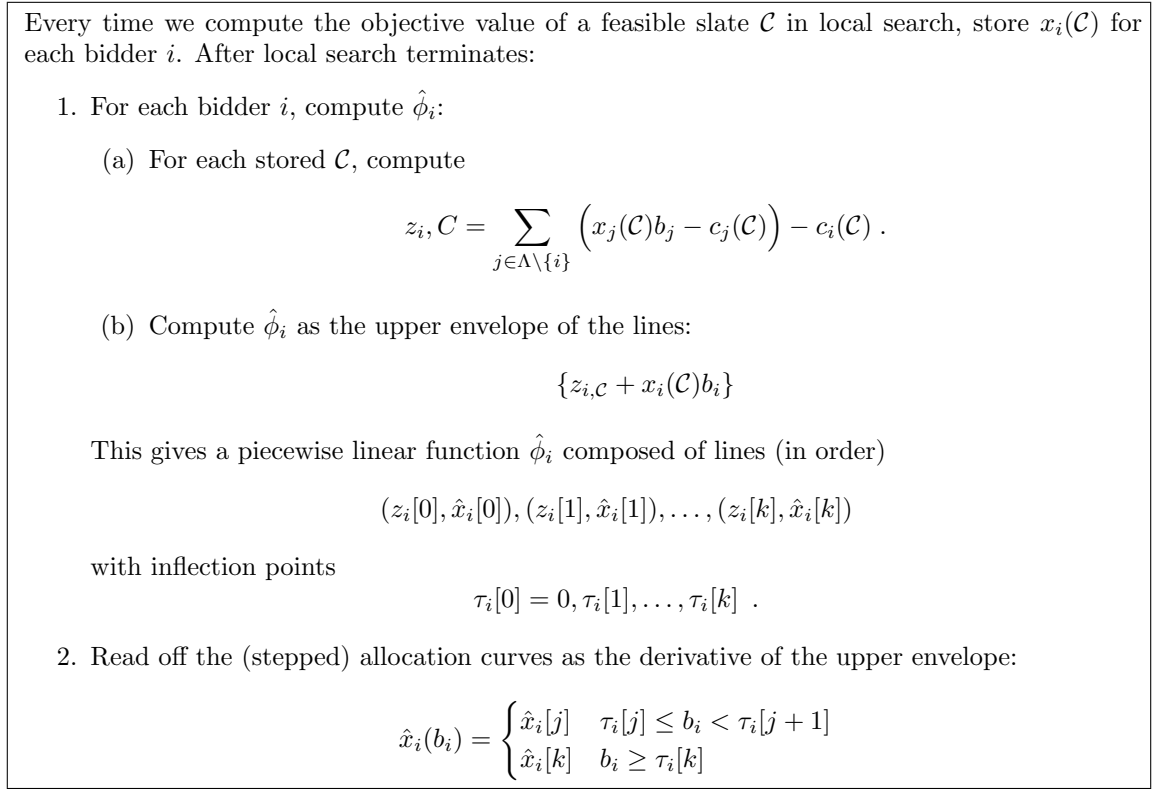

\center{ \fbox{ \begin{minipage}{0.9\textwidth}
Every time we compute the objective value of a feasible slate $\conf$
in local search, store $x_i(\conf)$ for each bidder $i$. After local search
terminates:
\begin{enumerate}
\item[1.] For each bidder $i$, compute $\hat\phi_i$:
\begin{enumerate}
\item For each stored $\conf$, compute
\[  z_i,C = \sum_{j \in \Lambda \setminus\{i\}} \Big( x_j(\conf) b_j -
c_j(\conf) \Big) - c_i(\conf)\;.  \]
\item Compute $\hat\phi_i$ as the upper envelope of the lines:
\[\{z_{i,\conf}+x_i(\conf)b_i\}\]
\end{enumerate} This gives a piecewise linear function $\hat\phi_i$ composed of
lines (in order) \[(z_i[0],\hat x_i[0]),(z_i[1],\hat x_i[1]),\dots,(z_i[k],\hat
x_i[k])\] with inflection points
\[\tau_i[0]=0,\tau_i[1],\dots,\tau_i[k]\enspace.\]

\item[2.] Read off the (stepped) allocation curves as the derivative of the upper envelope:
\[\hat x_i(b_i)=\begin{cases}\hat x_i[j]&\tau_i[j]\leq b_i <\tau_i[j+1]\\ \hat x_i[k]&b_i\geq \tau_i[k]\end{cases}\]
\end{enumerate}
\end{minipage}}}
\caption{\label{fig:price} Algorithm for constructing an estimated allocation curve $\hat x_i$.}
\end{figure}

This implies a straightforward method to compute $\hat x_i$, illustrated in
Figure~\ref{fig:price}.

\begin{observation}
The upper envelope is convex, therefore its slope is nondecreasing and the
allocation $\hat x_i$ is nondecreasing.\footnote{This should not be confused
with a claim that $x_i$ is non-decreasing; if the local search fails to
consider the right set of possible slates, its suboptimalities may lead
to non-monotonicities in the actual allocation function $x$. $\hat{x}_i$
remains monotonic by construction.}
\end{observation}

\subsection{Pricing methodologies}

The beauty of an approach like this, which efficiently constructs an accurate
representation of an entire allocation curve for each bidder, is that an array
of diverse pricing functions can be accommodated---all with the same underlying
infrastructure---with only a quick slate of the final ``price
read-off'' stage (step 2 in Figure \ref{fig:price}).

While we will ultimately settle on prices that mimic GSP, three pricing
strategies are worthy of discussion here: first-pricing, VCG pricing, and GSP
pricing. Each strategy has its own strengths and weaknesses.

\paragraph{First-pricing}  \vspace{1mm}
First-price auctions (advertisers pay exactly what they bid) are convenient to
implement but create major issues. Simple implementations are proven to be
unstable both in theory and in practice~\cite{edelman07}. While stability can
be restored~\cite{hoy13}, bidders must adopt a new bidding language. Perhaps
more damningly, first-price semantics would likely upset advertisers who are
generally accustomed to a second-price discount on search.

\paragraph{VCG pricing}  \vspace{1mm}
Running a traditional Vickrey-Clarke-Groves (VCG) auction is appealing for many
reasons, but is ultimately an undesirable solution. On the plus side, first,
standard theory says that it is the truthful auction. Second, VCG prices can be
efficiently computed as externalities --- it is sufficient to rerun the
optimization as a black box $n$ additional times, then compute the negative
effect each bidder has on the others. This mathematical abstraction naturally
leads to a practical implementation abstraction, making VCG prices easy to
implement. As a result, VCG has become the industry standard auction when
facing a complex optimization problem~\cite{varian14}.

However, VCG is not a perfect solution. Practically speaking, the marketplaces
that use VCG pricing have generally done so from an early stage --- we are
unaware of any mature markets that have {\em transitioned} from GSP to VCG. The
main challenge is that advertisers will need to change their bidding
strategies; until they do, the auctioneer will generally lose money. Even
assuming bidders eventually react, obtaining a smooth transition is a tricky
task~\cite{bachrach15}. Even worse in our particular circumstance, it is
unclear that advertisers will be responsive given Yahoo's market share.

More subtly, it is not clear that VCG is truly the best auction from a
theoretical viewpoint for reasons having to do with questions regarding which
utility model best reflects advertiser preferences. In particular, our prior
work even suggests that GSP might be the appropriate incentive compatible
auction~\cite{cavallo15}.

\paragraph{Generalized GSP pricing}  \vspace{1mm}
A natural solution is to stay with GSP pricing; the challenge is to define what
that means. A traditional GSP auction sorts ads by a ranking score and charges
each bidder the minimum bid required to hold its rank. This is sensible when
the auction is simply assigning ads to ranks; but when the auction makes a
complex trade-off over the features of an ad, this is no longer well-defined. A
theoretical literature strives to justify GSP's use; however, it fails to
identify the defining properties of GSP that one would need in order to
generalize it.

Based on our prior work, we propose that GSP be generalized as the truthful
auction for value maximizing bidders (see~\cite{cavallo15,wilkens16} for a
thorough treatment). A value maximizing bidder wants to get as many clicks as
possible without paying more than its value, i.e., to maximize $x_i$ while
keeping $p_i\leq v_i$. In contrast, a traditional model assumes bidders
maximize expected profit $(v_i-p_i)x_i$.

Defining truthful prices for these
bidders in our auction leads to a pricing intuition often given to the GSP
price:\footnote{This property of GSP prices is not a new observation,
but~\cite{cavallo15} is the first to give a solid foundation for why this
property is significant.}

\begin{definition}
The {\em truthful price $p_i$ for a value maximizer} is $p_i[j]=\tau_i[j]$ when $i$ gets allocation $x_i[j]$.
\end{definition}
That is, $p_i$ is the minimum bid advertiser $i$ must submit to maintain the same allocation.

\begin{figure}[h!] 
\center \includegraphics[scale=0.8]{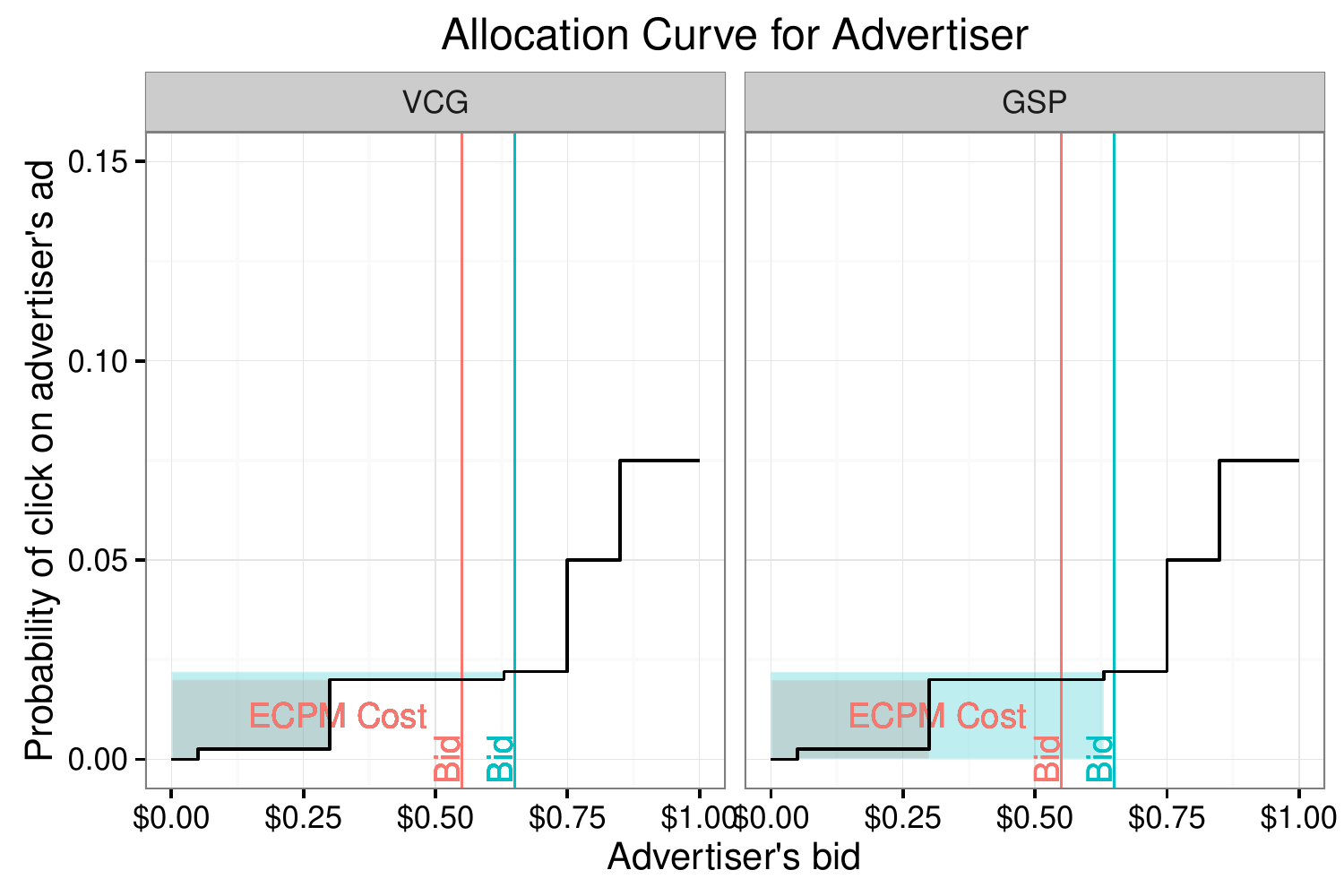} 
\caption{\label{fig:overcharge} An illustration of ``overcharging'' under generalized GSP payments. The red and blue shaded areas represent the prices charged for the red and blue bids respectively, under VCG (left) and generalized GSP (right) prices. Note that the small increase in click probability results in a small change in the VCG price but a large increase in the generalized GSP price.}
\end{figure}

This gives us a candidate auction: when $i$ gets allocation $x_i[j]$, charge $p_i[j]=\tau_i[j]$. Unfortunately, this
auction may ``overcharge.'' For example, if $x_i[j]\approx x_i[j-1]$ (there's a
tiny step in the allocation function) but $\tau_i[j]\geq 2\tau_i[j-1]$ (there's
a large difference in the minimum bids that yield the two allocations), an
advertiser might not care whether it gets allocation $x_i[j]$ or $x_i[j-1]$,
but this GSP auction could charge a 2x premium for the higher allocation. This
is illustrated in Figure~\ref{fig:overcharge}. The problem arises because the
value maximizing model assumes bidders are willing to pay an unrealistically
large price for a tiny increase in allocation.

To refine our version of GSP, we choose a middle ground between VCG pricing and
GSP pricing using ideas introduced in~\cite{cavallo15,wilkens16}. Our approach
is to start with a hybrid preference model --- a model of bidder preferences
that lies between quasilinear utilities and value maximizing preferences ---
and set prices so that bidders of the chosen type would be truthful. We propose two different hybrids.

\begin{figure}[h!] 
\center \includegraphics[scale=0.8]{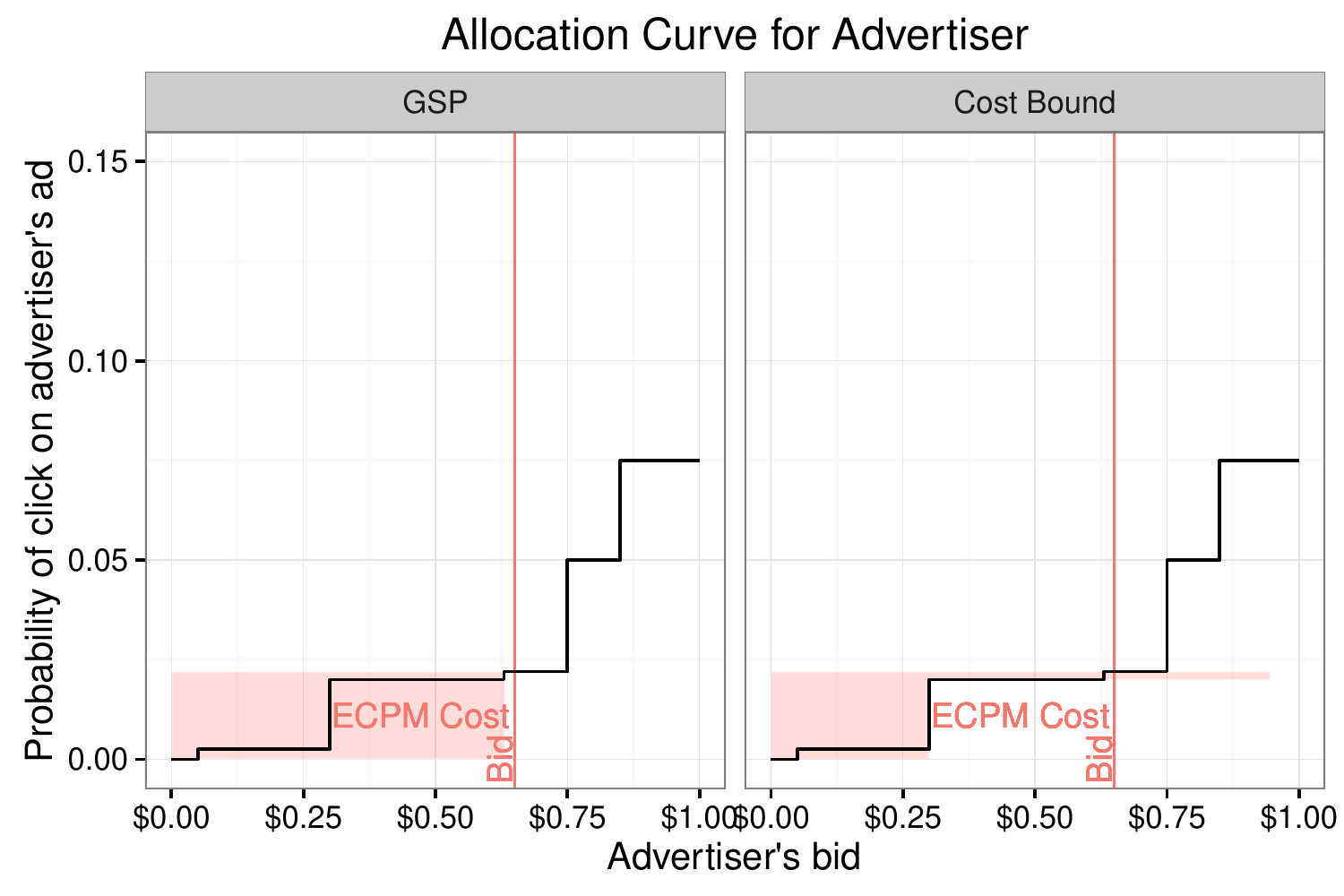} 
\caption{\label{fig:roi-price} An illustration of pricing for ROI-constrained bidders. Pricing for bidders with an ROI constraint charges the smaller of two quantities: the GSP price (left) and a maximum-marginal-cost-per-click increase over the previous price (right).}
\end{figure}

Our first hybrid model adds a return on investment (ROI) constraint of $\alpha$ to
existing quasilinear utilities. We refer the reader to~\cite{wilkens16} for
details, but the prices are as follows:

\begin{definition}
The {\em ROI-constrained truthful price} $p_i[j]$ when $i$ gets allocation $x_i[j]$ is computed by the following
recursive formula: $p_i[0]=0$, and for all $j > 0$,
\begin{align}
\hspace{-3mm} p_i[j] = \min\left(\tau_i[j], \;\frac{\hat x_i[j-1]p_i[j-1]+(\hat
x_i[j]-\hat x_i[j-1])(\alpha+1) \tau_i[j]}{\hat x_i[j]}\right)
\label{roiPrices}
\end{align}
\end{definition}
This formula has a natural interpretation: a bidder is charged the lesser of
the GSP price ($\tau_i[j]$) and the price one computes starting with $x_i[j-1]$
at price $p_i[j-1]$ and assuming a marginal cost-per-click of
$(\alpha+1)\tau_i[j]$ for the extra $x_i[j]-x_i[j-1]$ expected clicks. This is
illustrated in Figure~\ref{fig:roi-price}.

The second type of preferences we propose is continuous and assumes that
bidders optimize a utility function of the form
$u_i=v_i^{\alpha+1}-p_i^{\alpha+1}$. Again, we refer the reader
to~\cite{wilkens16} for details: \begin{definition}
The {\em $\alpha$-hybrid truthful price} $p_i[j]$ is given by the following
formula:
\begin{align}
p_i[j] = \frac{1}{x_i[j]}\left(\sum_{j=1}^i(\tau_i[j]\hat
x_i[j])^{\alpha+1}-(\tau_i[j]\hat
x_i[j-1])^{\alpha+1}\right)^{\frac1{\alpha+1}}  \label{alphaPrices}
\end{align}
\end{definition}
At $\alpha=0$, both models describe traditional VCG prices (truthful prices for
profit maximizers); as $\alpha\rightarrow\infty$, both models converge to GSP
prices (truthful for value maximizers). We choose a hybrid model to mimic GSP
while curtailing extremely high marginal prices.

\vspace{15mm}
\section{Results}  \label{sec:res}

\subsection{Allocation accuracy}

The first results we present regard how well our heuristic ad allocation
algorithm approximates the optimal solution.\footnote{The optimal allocation,
given any set of ad candidates, can be solved by a variety of methods including
integer programming; it is easy for us to compute statistics about an optimal
algorithm offline, despite it not being suitable for online use due to the
severe runtime constraints of our domain.}
We report results on a random selection of 100,000 auction instances drawn from
Yahoo's Gemini search platform for Desktop devices. An ``instance'' consists of
a set of candidate ads and all relevant accompanying information (bids,
clickability predictions, size, decorations, etc.).
We ran the algorithm with a variety of different maximum ad cardinality limits
(\texttt{ADLIM}), in each case applying a maximum number of ad lines
(\texttt{H}) equal to 18.

\begin{table}[h!]
\begin{center}
\begin{tabular}{c|c|c|c|c}
\texttt{ADLIM}& 2& 3& 4& 5 \\
\hline
Efficiency rate& 0.9999& 0.9998& 0.9978& 0.9957 \\
Optimality rate& 0.999& 0.993& 0.877& 0.806 \\
\end{tabular}
\caption{\label{tab:eff} Performance comparison of our heuristic algorithm against the optimal algorithm. The {\em efficiency rate} is the average ratio of our algorithm's efficiency to that of the optimal algorithm; the {\em optimality rate} is the percentage of instances on which our heuristic returned a globally optimal solution (i.e., achieved {\em efficiency rate} of 1).}
\end{center}
\end{table}

When a maximum of two ads may be shown, the heuristic misses the optimal
allocation in less than one out of every 10,000 instances. When three ads may
be shown, this goes down to about one out of every 150 instances. As the
\texttt{ADLIM} increases the heuristic diverges from the optimal solution in
more and more cases; however, when it does diverge, it still finds a solution
that is negligibly worse than the optimal one. Even for an \texttt{ADLIM} of 5
(which is the upper limit of what is currently seen on any of the major search
platforms), our heuristic algorithm obtains more than 99.5\% of the optimal
efficiency on average.

\subsection{Pricing accuracy}

Completely apart from the potential suboptimality of the allocation that our
algorithm computes, there is approximation in the prices we compute. As
discussed, to precisely compute prices (say, according to Eq.~(\ref{roiPrices})
or Eq.~(\ref{alphaPrices})) one needs to compute the {\em allocation curve} for
the bidder, from which the price can be quickly deduced. One can do so in a
brute-force manner, panning across the space of possible bids and observing how
the bidder's allocation (and predicted number of clicks) changes, holding all
other bids constant. Since there are only a finite number of allocations, one
can do better than this by using binary search to determine the ``break
points'' in the allocation curve---i.e., the set of distinct flat regions it is
constituted by. But even this will be too computationally costly to use in
real-time. Hence our {\em online} method for computing allocation curves,
described in Section \ref{sec:pricing}.

\begin{figure}[h!] 
\center \includegraphics[width=0.7\textwidth]{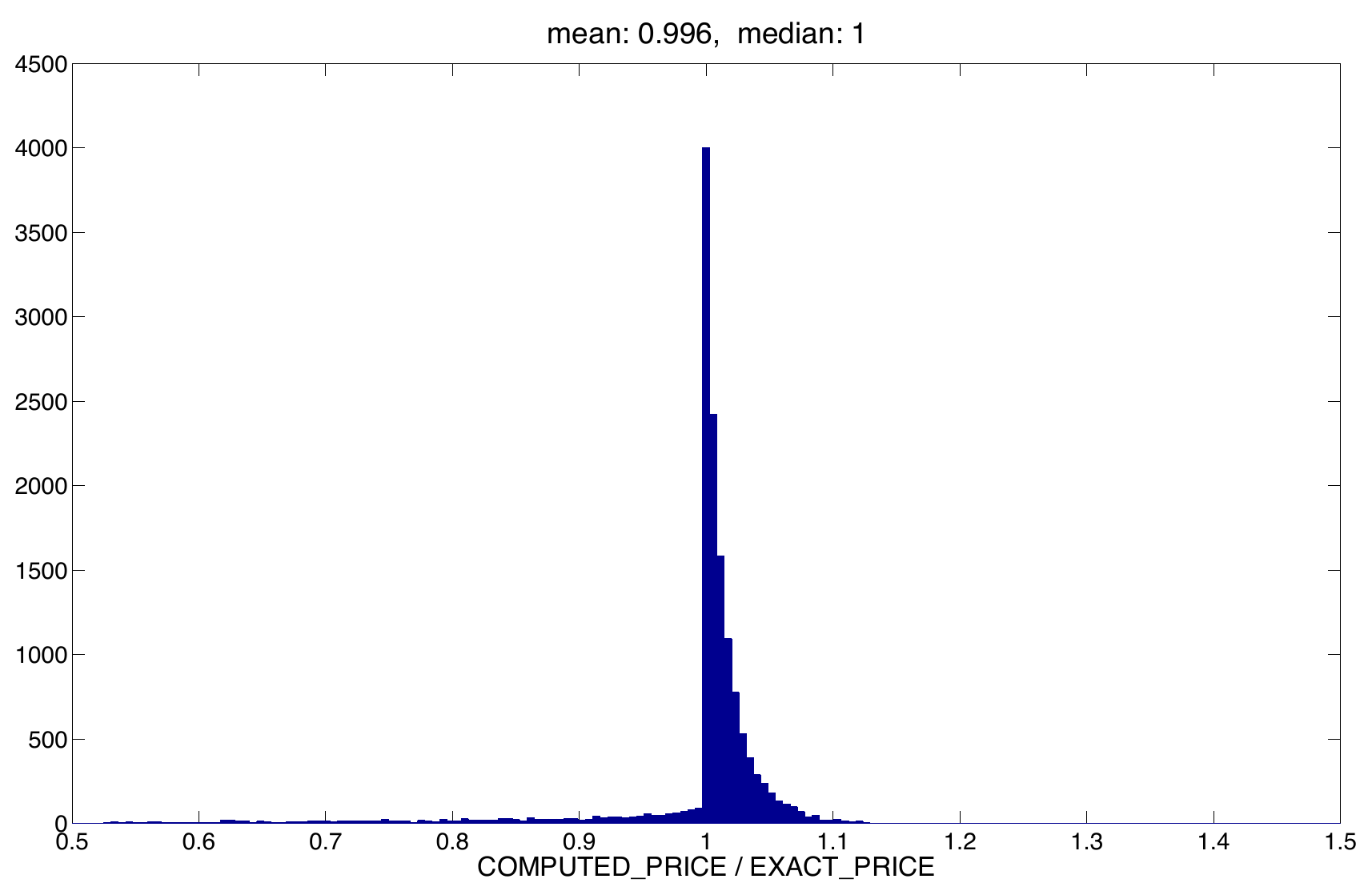} 
\center \includegraphics[width=0.7\textwidth]{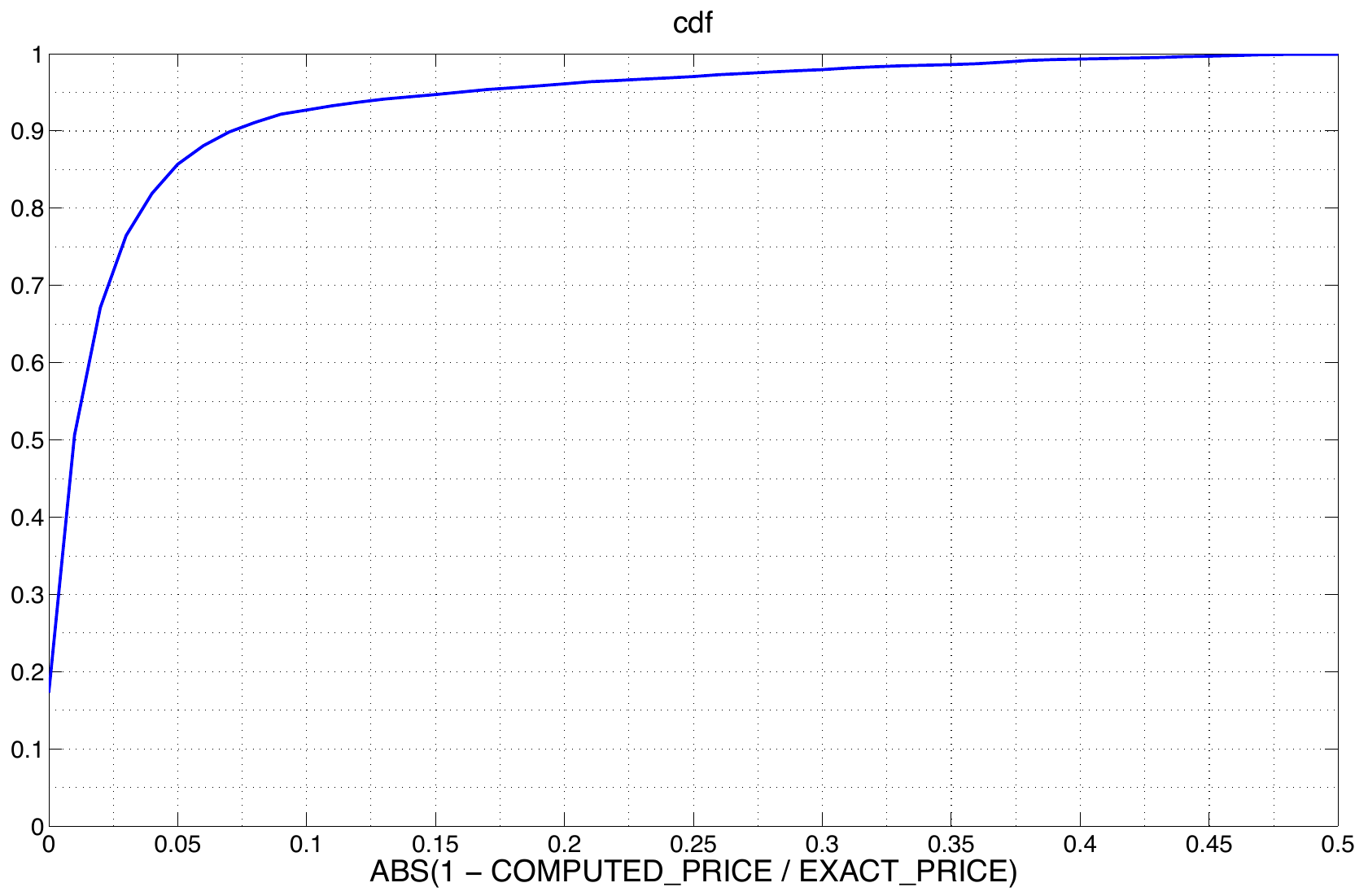} 
\caption{\label{fig:priceAcc} The distribution of the ratio of our computed prices to exact prices (top), and the cdf of the distance between our estimate and the exact price (bottom).}
\end{figure}

Determining those prices is computationally feasible, but are the prices any
good? Yes. Figure \ref{fig:priceAcc} illustrates the accuracy of our
``approximate prices'' by comparing them to the exact prices, computed offline
via binary search.

The top illustration in Figure \ref{fig:priceAcc} is a histogram of the ratios
of approximated price to exact price. The giant peak at 1 indicates that we get
the price precisely right a large percentage of the time. There is a
significant volume just to the right of peak, where we slightly overestimate
prices, and there is a fairly long and light tail below 1. The bottom
illustration conveys the cumulative distribution of the absolute-value distance
of our approximate prices from the exact prices.  We're within 5\% of the exact
price about 85\% of the time, and we're within 15\% almost 95\% of the time.

\subsection{Online bucket results}

In live bucket tests our algorithm substantially improves over its predecessor,
packing more ads into less space. Selected results from one test are given in
Table~\ref{tab:bucket}:  the test packed $>10\%$ more ads into $<90\%$ of the
lines. Meanwhile, it increased advertiser value by 8\% (assuming bids are
truthful) with clicks remaining nearly neutral. Revenue was also neutral, but
this metric has little meaning in a small bucket test when the auction rules
change.

\begin{table}[h]
\begin{center}
\begin{tabular}{l|r}
{\bf Metric}& {\bf Test vs. Control} \\
\hline
Total Ads& $+11\%$ \\ 
Total Lines& $-11\%$ \\ 
Value per Search& $+8\%$ \\ 
Revenue per Search& neutral\\ 
Click Yield& $-1\%$ 
\end{tabular}
\caption{\label{tab:bucket}Selected results from a bucket test on Yahoo's desktop search platform. The test ran for ~5 days on 1.5\% of traffic.}
\end{center}
\end{table}

\section{Discussion}  \label{sec:dis}

In this paper we provided a formal introduction of the {\em rich ads} problem
for sponsored search, and described our recent efforts to address it. The major
part of the paper focused on presenting the details of our engineered solution.
We described a local search based heuristic method that achieves performance
that is practically identical to that of an optimal algorithm, while meeting
the tight runtime constraints of the sponsored search domain. We described a
method that couples the algorithmic determination of a near-optimal allocation
with allocation-curve construction, which allows us to quickly compute prices
without repeating work, making the whole system runtime feasible and easily
adaptable to future developments.

The claim about our heuristic optimization algorithm being near-optimal is an
empirical observation based on the types of decorations available today and the
ad real-estate constraint in place. Similarly we have empirically established
the close approximation of our allocation curve generation method.  In our
future work, we will be exploring exact optimization algorithms that are
guaranteed to produce allocation curves with approximation bounds.

Online experiments on a fraction of Yahoo search traffic comparing the
performance of our algorithm with the standard GSP algorithm on metrics such as
revenue, click yield, ad real estate footprint, and user response indicate that
our new approach yields improved outcomes in a ``win-win-win'' fashion,
achieving gains in advertiser value and revenue, while at the same time
reducing the overall ad footprint.
\if We are currently running online experiments to compare the performance of
our algorithm with the standard GSP algorithm on metrics such as revenue, click
yield, ad real estate footprint and user response metrics.  While the
experiments are ongoing, we \fi
One thing we observe is that, for many ads, after a certain point the
click-through-rate (CTR) per line has diminishing returns and thus smaller ads
have a higher average click-through-rate per line.  Our algorithm therefore
often favors smaller ad variants over larger ones, packing more ads for the
same total number of lines on a page.
\if This is especially the case for ads that include non-footprint increasing
decorations such as icons, badges, etc. \fi

After a version of our algorithm is launched into full-scale production we
expect that advertisers will adjust their bids in an effort to have their
preferred ad variant appear.  While in our present version we assume that we
can drop decorations willy-nilly to vary the size of the ad, advertiser
preferences over their ad variants may present practical constraints.  It may
be necessary to allow advertisers to bid separately for each ad variant, so
that their preferences can be properly expressed.  This would raise a number of
challenges, among other things forcing us to modify how we generate allocation
curves for pricing.  It is an area that we will be studying further.

\bibliographystyle{plain} \bibliography{searchadopt}

\begin{thebibliography}{10}

\bibitem{bachrach15}
Yoram Bachrach, Sofia Ceppi, Ian~A. Kash, Peter Key, and Mohammad~Reza Khani.
\newblock Mechanism design for mixed ads.
\newblock In {\em Ad Auctions Workshop}, January 2015.

\bibitem{bachrach14}
Yoram Bachrach, Sofia Ceppi, Ian~A. Kash, Peter Key, and David Kurokawa.
\newblock Optimising trade-offs among stakeholders in ad auctions.
\newblock In {\em Proceedings of the 15th ACM Conference on Economics and
  Computation}, pages 75--92, 2014.

\bibitem{cavallo15}
Ruggiero Cavallo, Prabhakar Krishnamurthy, and Christopher~A. Wilkens.
\newblock On the truthfulness of gsp.
\newblock In {\em Eleventh Workshop on Sponsored Search Auctions}, 2015.

\bibitem{cavallo14}
Ruggiero Cavallo and Christopher~A. Wilkens.
\newblock {\em Web and Internet Economics: 10th International Conference, WINE
  2014, Beijing, China, December 14-17, 2014. Proceedings}, chapter GSP with
  General Independent Click-through-Rates, pages 400--416.
\newblock Springer International Publishing, Cham, 2014.

\bibitem{deng10}
Xiaotie Deng, Yang Sun, Ming Yin, and Yunhong Zhou.
\newblock {\em Mechanism Design for Multi-slot Ads Auction in Sponsored Search
  Markets}, pages 11--22.
\newblock Springer Berlin Heidelberg, Berlin, Heidelberg, 2010.

\bibitem{dutting16}
Paul D\"{u}tting, Felix Fischer, and David~C. Parkes.
\newblock Truthful outcomes from non-truthful position auctions.
\newblock In {\em Proceedings of the 2016 ACM Conference on Economics and
  Computation}, EC '16, pages 813--813, New York, NY, USA, 2016. ACM.

\bibitem{edelman07}
Benjamin Edelman and Michael Ostrovsky.
\newblock Strategic bidder behavior in sponsored search auctions.
\newblock {\em Decis. Support Syst.}, 43(1):192--198, February 2007.

\bibitem{eos07}
Benjamin Edelman, Michael Ostrovsky, and Michael Schwarz.
\newblock Internet advertising and the generalized second-price auction:
  Selling billions of dollars worth of keywords.
\newblock {\em American Economic Review}, 97(1):242--259, 2007.

\bibitem{fujishima1999taming}
Yuzo Fujishima, Kevin Leyton-Brown, and Yoav Shoham.
\newblock Taming the computational complexity of combinatorial auctions:
  Optimal and approximate approaches.
\newblock In {\em IJCAI}, volume~99, pages 548--553. DTIC Document, 1999.

\bibitem{gunluk2005branch}
Oktay G{\"u}nl{\"u}k, L{\'a}szlo Lad{\'a}nyi, and Sven De~Vries.
\newblock A branch-and-price algorithm and new test problems for spectrum
  auctions.
\newblock {\em Management Science}, 51(3):391--406, 2005.

\bibitem{hoy13}
Darrell Hoy, Kamal Jain, and Christopher~A. Wilkens.
\newblock A dynamic axiomatic approach to first-price auctions.
\newblock In {\em Proceedings of the Fourteenth ACM Conference on Electronic
  Commerce}, EC '13, pages 583--584, New York, NY, USA, 2013. ACM.

\bibitem{lubin08}
Benjamin Lubin, Adam~I. Juda, Ruggiero Cavallo, S{\'e}bastien Lahaie, Jeffrey
  Shneidman, and David~C. Parkes.
\newblock {ICE}: An expressive iterative combinatorial exchange.
\newblock {\em Journal of Artificial Intelligence Research}, 33(1):33--77,
  2008.

\bibitem{milgrom10}
Paul Milgrom.
\newblock Simplified mechanisms with an application to sponsored-search
  auctions.
\newblock {\em Games and Economic Behavior}, 70(1):62--70, 2010.

\bibitem{varian07}
Hal~R. Varian.
\newblock Position auctions.
\newblock {\em International Journal of Industrial Organization},
  25:1163--1178, 2007.

\bibitem{varian14}
Hal~R. Varian and Christopher Harris.
\newblock The vcg auction in theory and practice.
\newblock {\em American Economic Review}, 104(5):442--45, 2014.

\bibitem{wilkens16}
Christopher~A. Wilkens, Ruggiero Cavallo, and Rad Niazadeh.
\newblock Mechanism design for value maximizers.
\newblock {\em CoRR}, abs/1607.04362, 2016.

\end{thebibliography}

\end{document}